\pdfoutput=1
\RequirePackage{ifpdf}
\ifpdf 
\documentclass[pdftex]{sigma}
\else
\documentclass{sigma}
\fi

\begin{document}

\allowdisplaybreaks

\renewcommand{\thefootnote}{$\star$}

\renewcommand{\PaperNumber}{077}

\FirstPageHeading

\ShortArticleName{Semi-Inf\/inite $q$-Boson System}

\ArticleName{Boundary Interactions
\\
for the Semi-Inf\/inite $\boldsymbol{q}$-Boson System\\
and Hyperoctahedral Hall--Littlewood Polynomials\footnote{This paper is a~contribution to the Special Issue
in honor of Anatol Kirillov and Tetsuji Miwa.
The full collection is available at \href{http://www.emis.de/journals/SIGMA/InfiniteAnalysis2013.html}{http://www.emis.de/journals/SIGMA/Inf\/initeAnalysis2013.html}}}

\Author{Jan Felipe VAN DIEJEN and Erdal EMSIZ}
\AuthorNameForHeading{J.F.~van Diejen and E.~Emsiz}

\Address{Facultad de Matem\'aticas, Pontificia Universidad Cat\'olica de Chile,\\
Casilla 306, Correo 22, Santiago, Chile}
\Email{\href{mailto:diejen@mat.puc.cl}{diejen@mat.puc.cl}, \href{mailto:eemsiz@mat.puc.cl}{eemsiz@mat.puc.cl}}

\ArticleDates{Received September 27, 2013, in f\/inal form November 26, 2013; Published online December 04, 2013}

\vspace{-1mm}

\Abstract{We present a~semi-inf\/inite $q$-boson system endowed with a~four-parameter boun\-dary interaction.
The $n$-particle Hamiltonian is diagonalized by generalized Hall--Littlewood polynomials with
hyperoctahedral symmetry that arise as a~degeneration of the Macdonald--Koornwinder polynomials and were
recently studied in detail by Venkateswaran.}


\Classification{33D52; 81T25; 81R50; 82B23}

\renewcommand{\thefootnote}{\arabic{footnote}} 
\setcounter{footnote}{0}

\vspace{-2mm}

\section{Introduction}

The $q$-boson model introduced by Bogoliubov et al.~\cite{bog-ize-kit:correlation} is a~quantum many body
system on the one-dimensional lattice built of particle creation and annihilation operators representing
the $q$-oscillator algebra (cf., e.g.,~\cite[Section~3.1]{maj:foundations}
and~\cite[Chapter~5]{kli-sch:quantum} and references therein for further background material concerning the
$q$-oscillator algebra and its representations).
The model in question can be seen as a~limiting case of a~more general quantum particle system arising as
a~$q$-deformation of the totally asymmetric simple exclusion process
($q$-TASEP)~\cite{bor-cor-pet-sas:spectral,pov:integrability, sas-was:exact,tak:discrete}.
The $n$-particle Bethe ansatz eigenfunctions of the $q$-boson model amount to Hall--Littlewood polynomials,
both in the case of a~f\/inite periodic lattice (with f\/inite discrete
spectrum)~\mbox{\cite{kor:cylindric, tsi:quantum}} and in that of an inf\/inite lattice (with bounded absolutely
continuous spectrum)~\cite{die-ems:diagonalization}.
For appropriate boundary f\/ields acting on the particles at the end point of the semi-inf\/inite
lattice~\cite{die-ems:semi-infinite}, the Bethe ansatz eigenfunctions result moreover to be given by
Macdonald's three-parameter Hall--Littlewood polynomials with hyperoctahedral symmetry associated with the
root system~$BC_n$~\mbox{\cite[\S~10]{mac:orthogonal}}.

Recently it was pointed out that the $BC_n$-type Hall--Littlewood polynomials of Macdonald can be viewed as
a~subfamily of a~more general f\/ive-parameter family of hyperoctahedral Hall--Littlewood polynomials that
was studied in detail by Venkateswaran~\cite{ven:symmetric}; this f\/ive-parameter family arises as a~$q\to
0$ degeneration~-- without parameter conf\/luences~-- of the Macdonald--Koornwinder multivariate
Askey--Wilson polynomials~\cite{koo:askey-wilson,mac:affine}.
The purpose of the present note is to show that the f\/ive-parameter hyperoctahedral Hall--Littlewood
polynomials at issue constitute the eigenfunctions of a~semi-inf\/inite $q$-boson model endowed with
boundary interactions that involve both the particles at the end point of the lattice and those at its
nearest neighboring site.
The underlying boundary deformation of the $q$-boson f\/ield algebra violates the principle of
ultralocality: the particle creation and annihilation operators belonging to the end point and its nearest
neighboring site no longer commute, and moreover, the $n$-particle eigenfunctions are only of the usual
coordinate Bethe ansatz form away from the end point.
\begin{remark}
To avoid possible confusion, it is important to emphasize that the parameter $q$ of the $q$-boson model
does {\em not} correspond to the $q$-deformation parameter that enters in Macdonald's theory of orthogonal
polynomials associated with root systems~\cite{mac:orthogonal,mac:affine} but rather to the parameter $t$
used there.
A dif\/ferent parameter $t$ is employed below to abbreviate our notation for a~frequently appearing product
comprised by the four Askey--Wilson-type parameters $t_1,\ldots, t_4$ of the Macdonald--Koornwinder
polynomial (and its ($q\to 0$) Hall--Littlewood-type degeneration).
\end{remark}

\section[Hyperoctahedral Hall-Littlewood polynomials]{Hyperoctahedral Hall--Littlewood polynomials}

\subsection{Orthogonality}
\label{orthogonality:sec}
Let $W$ be the hyperoctahedral group formed by the semi-direct product of the symmetric group~$S_n$ and the
$n$-fold product of the cyclic group $\mathbb{Z}_2\cong \{1,-1\}$.
An element \mbox{$w=(\sigma,\epsilon)\in W$} acts naturally on $\xi=(\xi_1,\ldots,\xi_n)\in \mathbb{R}^n$ via
$w\xi:=(\epsilon_1 \xi_{\sigma_1},\ldots,\epsilon_n \xi_{\sigma_n})$ (with $\sigma\in S_n$ and
$\epsilon_j\in \{1,-1\}$ for $j=1,\ldots,n$).
The algebra $A$ of $W$-invariant polynomials on the torus $\mathbb{T}_n:=\mathbb{R}^n/(2\pi\mathbb{Z}^n)$
is spanned by the hyperoctahedral monomial symmetric functions
\begin{gather*}
m_\lambda(\xi)=\sum_{\mu\in W\lambda}e^{i\langle\mu,\xi\rangle},
\qquad
\lambda\in\Lambda_n,
\end{gather*}
where $\Lambda_n$ stands for the set of partitions $\lambda=(\lambda_1,\ldots,\lambda_n)\in\mathbb{Z}^n$
with the convention $\lambda_1\geq\dots\geq\lambda_n\geq 0$, and the summation is meant over the orbit of
$\lambda$ with respect to the action of $W$; the bracket $\langle\cdot,\cdot \rangle$ refers to the
standard inner product on $\mathbb{R}^n$, i.e.~$\langle \mu,\xi\rangle =\mu_1\xi_1+\dots +\mu_n\xi_n$.

The basis of hyperoctahedral Hall--Littlewood polynomials $\text{p}_\lambda(\xi)$, $\lambda\in\Lambda_n$
studied in~\cite{ven:symmetric} arises from the monomial basis via a~(partial) Gram--Schmidt-like process as the
trigonometric polynomials of the form
\begin{subequations}\label{hl}
\begin{gather}
\label{hl1}
\text{p}_\lambda(\xi)=m_\lambda(\xi)+\sum_{\substack{\mu\in\Lambda_n
\\
\text{with}\  \mu<\lambda}}c_{\lambda,\mu}m_\mu(\xi),
\qquad
 c_{\lambda,\mu}\in\mathbb{C},
\end{gather}
such that
\begin{gather}
\label{hl2}
\langle\text{p}_\lambda,m_\mu\rangle_\Delta=0
\qquad
\text{if}\quad\mu<\lambda
\end{gather}
\end{subequations}
(so $\langle \text{p}_\lambda,\text{p}_\mu\rangle_\Delta = 0$ if $\mu <\lambda $).
Here we have employed the hyperoctahedral dominance partial ordering of the partitions
\begin{gather}
\label{porder}
\forall\, \mu,\lambda\in\Lambda_n:
\quad
\mu\leq\lambda
\qquad
\text{if\/f}
\qquad
\sum_{1\leq j\leq k}\mu_j\leq\sum_{1\leq j\leq k}\lambda_j
\qquad
\text{for}
\quad
k=1,\ldots,n
\end{gather}
(which dif\/fers from the usual dominance partial order in that one does not demand the additional degree
homogeneity condition $\mu_1+\dots+\mu_n=\lambda_1+\dots +\lambda_n$ for the partitions to be comparable)
together with the following inner product on $A$:
\begin{subequations}
\begin{gather}
\langle f,g\rangle_\Delta:=\frac{1}{(2\pi)^n|W|}\int_{\mathbb{T}_n}f(\xi)\overline{g(\xi)}
|\Delta(\xi)|^2\text{d}\xi,
\qquad
f,g\in A,
\end{gather}
with $|W|=2^n n!$ denoting the order of the hyperoctahedral group and
\begin{gather}
\label{measure}
\Delta(\xi):=\prod_{1\leq j<k\leq n}\frac{\big(1-e^{i(\xi_j-\xi_k)}\big)\big(1-e^{i(\xi_j+\xi_k)}\big)}
{\big(1-q e^{i(\xi_j-\xi_k)}\big)\big(1-q e^{i(\xi_j+\xi_k)}\big)}\prod_{1\leq j\leq n}
\frac{1-e^{2i\xi_j}}{\prod\limits_{r=1}^4\big(1-t_r e^{i\xi_j}\big)}.
\end{gather}
\end{subequations}
Throughout it is assumed that the parameters belong to the domain
\begin{gather*}
q\in(0,1)
\qquad
\text{and}
\qquad
t_r\in(-1,1)\setminus\{0\},
\qquad
 r=1,\ldots,4.
\end{gather*}

The hyperoctahedral Hall--Littlewood polynomials satisfy the following orthogonality
rela\-tions~\cite{ven:symmetric}:
\begin{subequations}\label{ortho-norm}
\begin{gather}\label{ortho}
\langle\text{p}_\lambda,\text{p}_\mu\rangle_\Delta=
\begin{cases}
0&\text{if}\quad\lambda\neq\mu,
\\
\mathcal{N}_\lambda&\text{if}\quad\lambda=\mu,
\end{cases}
\end{gather}
where
\begin{gather}
\label{norm}
\mathcal{N}_\lambda
:=\frac{(1-q)^{n}\big(tq^{m_0(\lambda)-1}\big)_{m_0(\lambda)}}
{(tq^{2m_0(\lambda)})_{m_1(\lambda)}\prod\limits_{1\leq r<s\leq4}(t_rt_s)_{m_0(\lambda)}\prod\limits_{l\geq0}(q)_{m_l(\lambda)}}
\qquad
\text{with}
\quad
t:=t_1t_2t_3t_4.
\end{gather}
\end{subequations}
Here the multiplicity $m_l(\lambda)$ counts the number of parts $\lambda_j$, $1\leq j\leq n$ of size
$\lambda_j=l$ (so $m_0(\lambda)$ is equal to $n$ minus the number of nonzero parts) and we have used
$q$-shifted factorials
\begin{gather*}
(x)_m:=(1-x)(1-xq)\cdots\big(1-xq^{m-1}\big)
\end{gather*}
with the convention that $(x)_0=1$.
Notice that the orthogonality $\langle \text{p}_\lambda,\text{p}_\mu\rangle_\Delta=0$ for distinct
partitions $\lambda$ and $\mu$ is manifest from the def\/ining properties in
equations~\eqref{hl} when both weights are {\em comparable} in the hyperoctahedral dominance
partial ordering~\eqref{porder}, whereas for noncomparable partitions the orthogonality is not at all
obvious from the above construction.

\subsection{Explicit formula}

The orthogonality relations in equations~\eqref{ortho-norm}~-- which
arise as a~($q\to 0$) degeneration of well-known orthogonality relations for the Macdonald--Koornwinder
multivariate Askey--Wilson polynomials~\cite{die:properties,koo:askey-wilson,mac:affine}~--
form a~two-parameter extension of Macdonald's orthogonality relations for the Hall--Littlewood polynomials
associated with the root system $BC_n$~\cite[\S~10]{mac:orthogonal}.
An explicit formula for the hyperoctahedral Hall--Littlewood polynomials~\eqref{hl}
generalizing the corresponding classic formula of Macdonald is given by~\cite{ven:symmetric}
\begin{subequations}\label{HL-monic-normalization}
\begin{gather}
\label{HL}
\text{p}_\lambda(\xi)=\frac{1}{n_\lambda}\sum_{w\in W}C_\lambda(w\xi)e^{-i\langle\lambda,w\xi\rangle},
\end{gather}
with
\begin{gather}\label{Cf}
C_\lambda(\xi):=\prod_{1\leq j<k\leq n}\frac{\big(1-q e^{i(\xi_j-\xi_k)}\big)\big(1-q e^{i(\xi_j+\xi_k)}\big)}
{\big(1-e^{i(\xi_j-\xi_k)}\big)\big(1-e^{i(\xi_j+\xi_k)}\big)}\prod_{\substack{1\leq j\leq n\\\lambda_j>0}}
\frac{\prod\limits_{r=1}^4\big(1-t_r e^{i\xi_j}\big)}{1-e^{2i\xi_j}}
\end{gather}
and
\begin{gather}
\label{monic-normalization}
n_\lambda:=(1-q)^{-n}(-1)_{m_0(\lambda)}\big(t q^{2m_0(\lambda)}\big)_{m_1(\lambda)}\prod_{l\geq0}(q)_{m_l(\lambda)}.
\end{gather}
\end{subequations}

\subsection{Pieri-type recurrence relation}

The ($q\to 0$) degeneration of a Pieri-type recurrence relation
for the Macdonald--Koornwinder multivariate Askey--Wilson polynomials~\cite[Section~6]{die:properties}
readily entails a corresponding recurrence relation for the normalized hyperoctahedral Hall--Littlewood polynomials
\begin{subequations}\label{n-pol}
\begin{gather}
\label{n-pol1}
P_\lambda(\xi):=c_\lambda\text{p}_\lambda(\xi),
\end{gather}
where
\begin{gather}
\label{n-pol2}
c_\lambda:=\frac{\tau_1^{\lambda_1}\cdots\tau_n^{\lambda_n}\big(t q^{2m_0(\lambda)}\big)_{m_1(\lambda)}
\prod\limits_{l\geq0}(q)_{m_l(\lambda)}}{(q)_{n}\prod\limits_{1<r\leq4}(t_1t_rq^{m_0(\lambda)})_{n-m_0(\lambda)}}
\end{gather}
\end{subequations}
with $\tau_j:=q^{n-j}t_1$ for $j=1,\ldots,n$.
\begin{proposition}
[Pieri formula]
\label{pieri:prp}
The normalized hyperoctahedral Hall--Littlewood polynomials $P_\lambda (\xi)$, $\lambda\in\Lambda_n$
satisfy the recurrence relation
\begin{gather}
 P_\lambda(\xi)\sum_{j=1}^n\big(2\cos(\xi_j)-\tau_j-\tau_j^{-1}\big) \nonumber
\\
\qquad{} =
\sum_{\substack{1\leq j\leq n\\
\text{\rm s.t.} \
\lambda+e_j\in\Lambda_n}}V_j^+(\lambda)\left(P_{\lambda+e_j}
(\xi)-P_\lambda(\xi)\right)+\sum_{\substack{1\leq j\leq n\\\text{\rm s.t.} \
\lambda-e_j\in\Lambda_n}}V_j^-(\lambda)\left(P_{\lambda-e_j}
(\xi)-P_\lambda(\xi)\right),
\label{pieri}
\end{gather}
with the vectors $e_1,\ldots, e_n$ denoting the standard unit basis of $\mathbb{Z}^n$ and
\begin{gather*}
V_j^+(\lambda)
:=\tau_j^{-1}[m_{\lambda_j}(\lambda)]\big(1-t q^{2m_0(\lambda)
+m_1(\lambda)-1}\big)^{\delta_{\lambda_j-1}+\delta_{\lambda_j}}
\\
\phantom{V_j^+(\lambda):=}{}
\times\left(\frac{\prod\limits_{1<r\leq4}\big(1-t_1t_rq^{m_0(\lambda)-1}\big)}{\big(1-t q^{2m_0(\lambda)-2}
\big)\big(1-t q^{2m_0(\lambda)-1}\big)}\right)^{\delta_{\lambda_j}},
\\
V_j^-(\lambda)
:=\tau_j[m_{\lambda_j}(\lambda)]\left(\frac{(1-t q^{m_0(\lambda)-1})\prod\limits_{1<r<s\leq4}
\big(1-t_rt_sq^{m_0(\lambda)}\big)}{\big(1-t q^{2m_0(\lambda)-1}\big)\big(1-t q^{2m_0(\lambda)}\big)}\right)^{\delta_{\lambda_j-1}}.
\end{gather*}
Here we have employed the $q$-integers $[m]:=(1-q^m)/(1-q)$ for $m=0,1,2,3,\ldots$ as well as the discrete
delta function on $\mathbb{Z}$: $\delta_l:=1$ if $l=0$ and $\delta_l:=0$ otherwise $($and the abbreviation
`s.t.' in the conditional sums on the r.h.s.\ of the recurrence stands for `such that'$)$.
\end{proposition}

\begin{proof}
As a~($q\to 0$) degeneration of the principal specialization formula for the Macdonald--Koornwinder
polynomials (see, e.g.,~\cite[equations~(6.1), (6.18), (6.43a)]{die:properties}) one f\/inds that
(assu\-ming momentarily $t_1>0$):
\begin{gather*}
\text{p}_\lambda\left(i\log(\tau_1),\ldots,i\log(\tau_n)\right)=\frac{1}{c_\lambda},
\end{gather*}
with $c_\lambda$ taken from equation~\eqref{n-pol2}.
This implies that the normalization of $P_\lambda (\xi )$~\eqref{n-pol} is such that the
polynomials in question satisfy a~($q\to 0$) degeneration of the Pieri-type recurrence formula
in equations~(6.4), (6.5), (6.12), (6.13) of~\cite{die:properties}, which~-- upon performing the limit~--
produces equation~\eqref{pieri}.
\end{proof}

\section[Boundary interactions for the semi-inf\/inite $q$-boson system]{Boundary interactions for the
semi-inf\/inite $\boldsymbol{q}$-boson system}

\subsection[Deformed $q$-boson f\/ield algebra]{Deformed $\boldsymbol{q}$-boson f\/ield algebra}
\label{bc-q-boson}
Let $\ell^2(\Lambda_n,\mathcal{N})$ be the Hilbert space of functions $f:\Lambda_n\to\mathbb{C}$ determined
by the inner product
\begin{gather*}
\langle f,g\rangle_n:=\sum_{\lambda\in\Lambda_n}f(\lambda)\overline{g(\lambda)}\mathcal{N}_\lambda,
\qquad
 f,g\in\ell^2(\Lambda_n,\mathcal{N}) ,
\end{gather*}
with $\mathcal{N}_\lambda$ given by equation~\eqref{norm} and the convention that
$\Lambda_0:=\{\varnothing\}$ and $\ell^2(\Lambda_0,\mathcal{N}):=\mathbb{C}$.
We think of $\ell^2(\Lambda_n,\mathcal{N})$ as the Hilbert space for a~system of $n$ quantum particles on
the nonnegative integer lattice $\mathbb{N}:=\{ 0,1,2,\ldots\}$ (i.e.\
the parts $\lambda_j$, $j=1,\ldots, n$ of $\lambda\in\Lambda_n$ encode the positions of the particles in
question).
In the Fock space
\begin{gather}
\label{f-space}
\mathcal{H}:=\bigoplus_{n\geq0}\ell^2(\Lambda_n,\mathcal{N}),
\end{gather}
consisting of all sequences $\sum\limits_{n\geq 0} f_n$ with $f_n\in \ell^2(\Lambda_n,\mathcal{N})$ such
that $\sum\limits_{n\geq 0} \langle f_n,f_n\rangle_n<\infty$, we introduce bounded annihilation operators
$\beta_l$, $l\in\mathbb{N}$ that are perturbed at the boundary site $\ell =0$ and act on $f\in
\ell^2(\Lambda_n,\mathcal{N})$ via
\begin{subequations}\label{an-op-cr-op}
\begin{gather}
\label{an-op}
(\beta_lf)(\lambda):=\frac{f(\beta_l^*\lambda)}{\big(1-tq^{2m_0(\lambda)+m_1(\lambda)}\big)^{\delta_l}},
\qquad
 \lambda\in\Lambda_{n-1},
\end{gather}
if $n>0$, and $\beta_lf:=0$ if $n=0$.
Here $\beta_l^*\lambda\in\Lambda_n$ is obtained from $\lambda$ by adding a~part of size $l$.
The action on $f\in \ell^2(\Lambda_n,\mathcal{N})$ of the adjoint of $\beta_l$ in $\mathcal{H}$ produces
the creation operator
\begin{gather}
\label{cr-op}
(\beta_l^*f)(\lambda)=f(\beta_l\lambda)[m_l(\lambda)]\big(1-tq^{2m_0(\lambda)+m_1(\lambda)-1}
\big)^{\delta_l+\delta_{l-1}}
\\
\phantom{(\beta_l^*f)(\lambda)=}
\times\left(\frac{\big(1-t q^{m_0(\lambda)-2}\big)\prod\limits_{1\leq r<s\leq4}\big(1-t_rt_sq^{m_0(\lambda)-1}\big)}
{\big(1-t q^{2m_0(\lambda)-3}\big)\big(1-t q^{2m_0(\lambda)-2}\big)^2\big(1-t q^{2m_0(\lambda)-1}\big)}\right)^{\delta_{l}},
\qquad
 \lambda\in\Lambda_{n+1},
\nonumber
\end{gather}
\end{subequations}
if $m_l(\lambda)>0$, and $(\beta_l^*f)(\lambda )=0$ otherwise.
Here $\beta_l\lambda\in\Lambda_n$ is obtained from $\lambda$ with $m_l(\lambda)>0$ by discarding a~part of
size $l$.
In the present setting, the role of the number operators is played by the bounded multiplication operators
\begin{gather}
\label{num-op}
(N_lf)(\lambda):=q^{m_l(\lambda)}f(\lambda),
\qquad
 f\in\ell^2(\Lambda_n,\mathcal{N}),\quad \lambda\in\Lambda_n,\quad l\in\mathbb{N} .
\end{gather}

When $t\neq q^m$ for $m=1,2,3,\ldots$, the creation and annihilation operators $\beta_l^*$, $\beta_l$
together with the commuting operators $N_l$, $(1-tq^cN_0^2)^{-1}$, $(1-tq^cN_0^2N_1)^{-1}$ (where
$l\in\mathbb{N}$ and $c\in\mathbb{Z}$) represent a~four-parameter deformation of the $q$-boson f\/ield
algebra at the boundary sites $l=0$ and $l=1$:
\begin{subequations}\label{com}
\begin{gather}
\label{com-a}
\beta_l N_k=q^{\delta_{l-k}}N_k\beta_l,
\qquad
\beta_l^*N_k=q^{-\delta_{l-k}}N_k\beta_l^*,
\\
\beta_l^*\beta_l=\frac{1-N_l}{1-q}\big(1-q^{-1}tN_0^2N_1\big)^{\delta_l+\delta_{l-1}}
\nonumber
\\
\phantom{\beta_l^*\beta_l=}{}
\times\left(\frac{\big(1-q^{-2}tN_0\big)\prod\limits_{1\leq r<s\leq4}\big(1-q^{-1}t_rt_sN_0\big)}{\big(1-q^{-3}
tN_0^2\big)\big(1-q^{-2}tN_0^2\big)^2\big(1-q^{-1}tN_0^2\big)\big(1-q^{-2}tN_0^2N_1\big)}\right)^{\delta_l},
\\
\beta_l\beta_l^*=\frac{1-qN_l}{1-q}\big(1-tN_0^2N_1\big)^{-\delta_l+\delta_{l-1}}\times{}
\nonumber
\\
\phantom{\beta_l\beta_l^*=}{}
\times\left(\frac{\big(1-q^{-1}tN_0\big)\big(1-qtN_0^2N_1\big)\prod\limits_{1\leq r<s\leq4}(1-t_rt_sN_0)}{\big(1-q^{-1}
tN_0^2\big)\big(1-tN_0^2\big)^2\big(1-qtN_0^2\big)}\right)^{\delta_l},
\end{gather}
and for $l<k$
\begin{gather}
\beta_l\beta_k=\left(\frac{1-qtN_0^2N_1}{1-tN_0^2N_1}\right)^{\delta_l\delta_{k-1}}\beta_k\beta_l,
\qquad
\beta_l^*\beta_k^*=\beta_k^*\beta_l^*\left(\frac{1-tN_0^2N_1}{1-qtN_0^2N_1}\right)^{\delta_l\delta_{k-1}},
\end{gather}
and
\begin{gather}
\label{com-e}
\beta_l\beta_k^*=\left(\frac{1-qtN_0^2N_1}{1-tN_0^2N_1}\right)^{\delta_l\delta_{k-1}}\beta_k^*\beta_l,
\qquad
\beta_l^*\beta_k=\beta_k\beta_l^*\left(\frac{1-tN_0^2N_1}{1-qtN_0^2N_1}\right)^{\delta_l\delta_{k-1}}.
\end{gather}
\end{subequations}
Indeed, it is straightforward to verify the commutation relations in equations~\eqref{com}
upon computing the explicit actions of both sides on an arbitrary function $f\in
\ell^2(\Lambda_n,\mathcal{N})$ with the aid of the formulas
in equations~\eqref{an-op-cr-op} and~\eqref{num-op}.

\subsection{Hamiltonian}
The Hamiltonian of our
semi-infinite $q$-boson system with boundary interaction is of the form
\begin{gather}
\label{hamiltonian}
H=V(N_0,N_1)+\sum_{l\in\mathbb{N}}\big(\beta_{l}^*\beta_{l+1}+\beta_{l+1}^*\beta_l\big),
\end{gather}
where $V(N_0,N_1)$ denotes a~boundary potential that depends rationally on $N_0$ and $N_1$.
By construction, $H$~\eqref{hamiltonian} preserves the $n$-particle sector
$\ell^2(\Lambda_n,\mathcal{N})\subset \mathcal{H}$ and we will denote the restriction of the Hamiltonian to
this $n$-particle subspace by~$H_n$.

\begin{proposition}[$n$-particle Hamiltonian]\label{Hn:prp}
For any $f\in \ell^2(\Lambda_n,\mathcal{N})$ and $\lambda\in\Lambda_n$, one has that
\begin{subequations}\label{Hn}
\begin{gather}
(H_{n}f)(\lambda)=V\big(q^{m_0(\lambda)},q^{m_1(\lambda)}\big)f(\lambda)
\nonumber
\\
\phantom{(H_{n}f)(\lambda)=}{}
+\sum_{\substack{1\leq j\leq n\\
\text{\rm s.t.}  \ \lambda+e_j\in\Lambda_n}}v_j^+(\lambda)f(\lambda+e_j)
+\sum_{\substack{1\leq j\leq n\\
\text{\rm s.t.} \ \lambda-e_j\in\Lambda_n}}v_j^-(\lambda)f(\lambda-e_j),
\label{Hna}
\end{gather}
with
\begin{gather}
v_j^+(\lambda):=[m_{\lambda_j}(\lambda)]\big(1-t q^{2m_0(\lambda)+m_1(\lambda)-1}\big)^{\delta_{\lambda_j-1}
+\delta_{\lambda_j}}
\nonumber
\\
\phantom{v_j^+(\lambda):=}{}
\times\left(\frac{\big(1-t q^{m_0(\lambda)-2}\big)\prod\limits_{1\leq r<s\leq4}\big(1-t_rt_sq^{m_0(\lambda)-1}\big)}
{\big(1-t q^{2m_0(\lambda)-3}\big)\big(1-t q^{2m_0(\lambda)-2}\big)^2\big(1-t q^{2m_0(\lambda)-1}\big)}\right)^{\delta_{\lambda_j}},
\label{Hnb}
\\
v_j^-(\lambda):=[m_{\lambda_j}(\lambda)].
\label{Hnc}
\end{gather}
\end{subequations}
\end{proposition}

\begin{proof}
It is immediate from the explicit actions of $\beta_l$ and $\beta_l^*$ in
equations~\eqref{an-op-cr-op} that for any $l\in\mathbb{N}$: $(\beta_{l+1} \beta_l^*
f)(\lambda)=0$ if $m_l(\lambda)=0$ and
\begin{gather*}
(\beta_l^*\beta_{l+1}f)(\lambda)
=[m_{\lambda_j}(\lambda)]\big(1-t q^{2m_0(\lambda)+m_1(\lambda)-1}\big)^{\delta_{l-1}+\delta_{l}}
\\
\phantom{(\beta_l^*\beta_{l+1}f)(\lambda)=}{}
\times\left(\frac{\big(1-t q^{m_0(\lambda)-2}\big)\prod\limits_{1\leq r<s\leq4}\big(1-t_rt_sq^{m_0(\lambda)-1}\big)}
{\big(1-t q^{2m_0(\lambda)-3}\big)\big(1-t q^{2m_0(\lambda)-2}\big)^2\big(1-t q^{2m_0(\lambda)-1}\big)}\right)^{\delta_{l}}
f(\beta_{l+1}^*\beta_l\lambda)
\end{gather*}
if $m_l(\lambda)>0$, where $\beta_{l+1}^* \beta_l\lambda=\lambda+e_j$ with $j=\min\{k\mid \lambda_k=l\}$
(so $l=\lambda_j$).
Along the same lines it is seen that $(\beta_{l+1}^*\beta_l f)(\lambda)=0$ if $m_{l+1}(\lambda)=0$ and
\begin{gather*}
(\beta_{l+1}^*\beta_l f)(\lambda)=[m_{l+1}(\lambda)]f(\beta_l^*\beta_{l+1}\lambda)
\end{gather*}
if $m_{l+1}(\lambda)>0$, where $\beta_l^*\beta_{l+1}\lambda=\lambda-e_j$ with $j=\max\{k\mid
\lambda_k=l+1\}$ (so $l=\lambda_j-1$).
The stated formula thus follows because the boundary potential acts (by def\/inition) via the
multiplication $(V(N_0,N_1)f)(\lambda)=V\big(q^{m_0(\lambda)},q^{m_1(\lambda)}\big)f(\lambda)$.
\end{proof}

\subsection{Diagonalization}

From now on we will pick the boundary potential $V(N_0,N_1)$ in~$H$ \eqref{hamiltonian} of the form
\begin{gather}
V(N_0,N_1)=
\left(t_1^{-1}tN_0+t_1N_0\left(1-\frac{\big(1-q^{-1}tN_0\big)\prod\limits_{1<r<s\leq4}(1-t_rt_s N_0)}{\big(1-tN_0^2\big)\big(1-q^{-1}
tN_0^2\big)}\right)\right)\frac{1-N_1}{1-q}\label{boundary-potential}
\\
\phantom{V(N_0,N_1)=}
{}+\left(t_1+qt_1^{-1}N_0^{-1}\left(1-\frac{\big(1-q^{-1}t N_0^2N_1\big)\prod\limits_{1<r\leq4}\big(1-q^{-1}t_1t_r N_0\big)}
{\big(1-q^{-2}tN_0^2\big)\big(1-q^{-1}tN_0^2\big)}\right)\right)\frac{1-N_0}{1-q}.
\nonumber
\end{gather}
By writing the action of $V(N_0,N_1)$~\eqref{boundary-potential} on an arbitrary $f\in \ell^2
(\Lambda_n,\mathcal{N})$ as a~rational expression in the parameters $t_r$ ($r=1,\ldots,4$), it is readily
seen~-- upon canceling possible common factors in the numerators and denominators~-- that $V(N_0,N_1)$
constitutes a~bounded multiplication operator in $\ell^2 (\Lambda_n,\mathcal{N})$.
It follows moreover from the Pieri recurrence in Proposition~\ref{pieri:prp} and the explicit formula for~$H_n$ in Proposition~\ref{Hn:prp} that the Hamiltonian with this boundary potential is diagonalized in the
$n$-particle subspace by a~hyperoctahedral Hall--Littlewood wave function $\phi_\xi:\Lambda_n\to\mathbb{C}$
of the form
\begin{gather}
\label{wave-function}
\phi_\xi(\lambda):=\frac{1}{\mathcal{N}_\lambda}\text{p}_\lambda(\xi),
\qquad
 \lambda\in\Lambda_n,
\end{gather}
where $\xi\in\mathbb{T}_n$ plays the role of the spectral parameter.

\begin{proposition}[$n$-particle eigenfunctions]\label{diagonal:prp}
The hyperoctahedral Hall--Littlewood wave func\-tion~$\phi_\xi$~\eqref{wave-function} satisfies the
eigenvalue equation
\begin{gather}
\label{ev-eq}
H_n\phi_\xi=E_n(\xi)\phi_\xi
\qquad
\text{with}
\quad
E_n(\xi):=2\sum_{1\leq j\leq n}\cos(\xi_j)
\end{gather}
for $H_n$ given by equations~\eqref{Hn} with $V(N_0,N_1)$ taken from
equation~\eqref{boundary-potential}.
\end{proposition}
\begin{proof}
By comparing the normalization of $\phi_\xi (\lambda)$~\eqref{wave-function}
and $P_\lambda(\xi)$~\eqref{n-pol},
one concludes that $\phi_\xi(\lambda)=\frac{1}{h_\lambda}P_\lambda (\xi)$ with
\begin{gather*}
h_\lambda=c_\lambda\mathcal{N}_\lambda
=\frac{\tau_1^{\lambda_1}\cdots\tau_n^{\lambda_n}\big(tq^{m_0(\lambda)-1}\big)_{m_0(\lambda)}
\prod\limits_{1<r<s\leq4}\big(t_rt_sq^{m_0(\lambda)}\big)_{n-m_0(\lambda)}}
{\big(tq^{n-1}\big)_{n}}\mathcal{N}_0.
\end{gather*}
It is thus immediate from equation~\eqref{pieri} that
\begin{gather*}
\begin{split}
& V\big(q^{m_0(\lambda)},q^{m_1(\lambda)}\big)\phi_\xi(\lambda)+\sum_{\substack{1\leq j\leq n\\
\text{\rm s.t.} \ \lambda+e_j\in\Lambda_n}}
v_j^+(\lambda)\phi_\xi(\lambda+e_j)+\sum_{\substack{1\leq j\leq n\\
\text{\rm s.t.} \  \lambda-e_j\in\Lambda_n}}
v_j^-(\lambda)\phi_\xi(\lambda-e_j)
\\
& \qquad
=E_n(\xi)\phi_\xi(\lambda),
\end{split}
\end{gather*}
with
\begin{gather*}
v_j^+(\lambda)=V_j^+(\lambda)\frac{h_{\lambda+e_j}}{h_\lambda},
\qquad
v_j^-(\lambda)=V_j^-(\lambda)\frac{h_{\lambda-e_j}}{h_\lambda}
\end{gather*}
and
\begin{gather*}
V\big(q^{m_0(\lambda)},q^{m_1(\lambda)}\big)=\sum_{1\leq j\leq n}\big(\tau_j+\tau_j^{-1}
\big)-\sum_{\substack{1\leq j\leq n
\\
\text{\rm s.t.} \ \lambda+e_j\in\Lambda_n}}V_j^+(\lambda)-\sum_{\substack{1\leq j\leq n
\\
\text{\rm s.t.} \ \lambda-e_j\in\Lambda_n}}V_j^-(\lambda).
\end{gather*}
By plugging in the explicit expressions for $V_j^+(\lambda)$, $V_j^-(\lambda)$, and $h_\lambda$, and
employing the elementary identity
\begin{gather*}
\sum_{1\leq j\leq n}\big(\tau_j+\tau_j^{-1}\big)-\sum_{\substack{1\leq j\leq n
\\
\text{\rm s.t.} \ \lambda+e_j\in\Lambda_n}}\tau_j^{-1}[m_{\lambda_j}
(\lambda)]-\sum_{\substack{1\leq j\leq n
\\
\text{\rm s.t.} \  \lambda-e_j\in\Lambda_n}}\tau_j[m_{\lambda_j}(\lambda)]=t_1[m_{0}(\lambda)],
\end{gather*}
the coef\/f\/icients $v_j^+(\lambda)$, $v_j^-(\lambda)$ and $V(q^{m_0(\lambda)},q^{m_1(\lambda)})$ are
rewritten in the form given by equations~\eqref{Hnb},~\eqref{Hnc} and~\eqref{boundary-potential}.
\end{proof}
\begin{remark}
The diagonalization in Proposition~\ref{diagonal:prp} in terms of the hyperoctahedral Hall--Little\-wood
polynomials implies that our $q$-boson Hamiltonian $H_n$ is unitarily equivalent to a~multiplication
operator governed by the eigenvalue~$E_n(\xi)$~\eqref{ev-eq}.
A complete system of commuting quantum integrals for~$H_n$ is obtained via this unitary equivalence from
the multiplication ope\-rators associated with the elements of the algebra $A$ of $W$-invariant trigonometric
polynomials on~$\mathbb{T}_n$.
It remains an open problem to present an explicit construction in the spirit
of~\cite{die-ems:diagonalization} that lifts~$H$~\eqref{hamiltonian} with~$V(N_0,N_1)$ given by
equation~\eqref{boundary-potential} to an inf\/inite hierarchy of commuting ope\-rators in the Fock space~$\mathcal{H}$~\eqref{f-space}, reproducing the quantum integrals of~$H_n$ upon restriction to the
$n$-particle subspace $\ell^2(\Lambda_n,\mathcal{N})$.
\end{remark}

\section{Ultralocality and coordinate Bethe ansatz}

For general parameter values the deformation of the $q$-boson field algebra in Section~\ref{bc-q-boson}
fails to be ultralocal, as the commutativity between the creation and
annihilation operators at sites $l=0$ and $l=1$ is broken.
The commutativity (and hence ultralocality) is restored when at least one of the four boundary parameters
$t_r$ tends to zero (so $t\to 0$).
It is furthermore clear from the explicit expression in equations~\eqref{HL-monic-normalization}
for the hyperoctahedral Hall--Littlewood polyno\-mial~$\text{p}_\lambda$~\eqref{hl} that the
wave function $\phi_\xi$~\eqref{wave-function} fails to be of the usual coordinate Bethe ansatz form (at
the boundary), as the expansion coef\/f\/icients $C_\lambda (w\xi)$ of the plane waves $e^{-i\langle
w\xi,\lambda\rangle}$ depend on (the number of nonzero parts of)~$\lambda$.
By letting at least two of the four boundary parameters $t_r$ tend to zero the polynomial
$\text{p}_\lambda$~\eqref{hl} reduces to Macdonald's Hall--Littlewood polynomial associated
with the root system of type $BC$, which implies that in this limiting case it is possible to rewrite the
wave function in the conventional Bethe ansatz form.
We end up by detailing our construction for these three- and two-parameter specializations of the boundary
interaction.

\subsection{Three-parameter reduction}

When $t_4\to 0$ (so $t\to 0$), the quadratic norm $\mathcal{N}_\lambda$~\eqref{norm} determining
inner product of the Fock space $\mathcal{H}$~\eqref{f-space} simplifies~to
\begin{gather*}
\mathcal{N}_\lambda=\frac{(1-q)^{n}}{\prod\limits_{1\leq r<s\leq3}(t_rt_s)_{m_0(\lambda)}\prod\limits_{l\geq0}
(q)_{m_l(\lambda)}}.
\end{gather*}
The actions of the annihilation and creation operators~\eqref{an-op-cr-op} on
$f\in\ell^2(\Lambda_n,\mathcal{N})$ then reduce to
\begin{gather*}
(\beta_lf)(\lambda)=f(\beta_l^*\lambda),
\qquad
 \lambda\in\Lambda_{n-1}
\end{gather*}
with the convention that $\beta_lf=0$ if $n=0$, and
\begin{gather*}
(\beta_l^*f)(\lambda)=f(\beta_l\lambda)[m_l(\lambda)]\prod_{1\leq r<s\leq3}
\big(1-t_rt_sq^{m_0(\lambda)-1}\big)^{\delta_l},
\qquad
 \lambda\in\Lambda_{n+1}
\end{gather*}
with the convention that $(\beta_l^*f)(\lambda )=0$ if $m_l(\lambda)=0$, respectively.
Together with the commuting operators $N_l$~\eqref{num-op} the creation and annihilation operators in
question represent a~three-parameter deformation of the $q$-boson f\/ield algebra at the boundary site~$l=0$:
\begin{gather*}
\beta_l N_k=q^{\delta_{l-k}}N_k\beta_l,
\qquad
\beta_l^*N_k=q^{-\delta_{l-k}}N_k\beta_l^*,
\\
\beta_l^*\beta_l=\frac{1-N_l}{1-q}\prod_{1\leq r<s\leq3}\big(1-q^{-1}t_rt_sN_0\big)^{\delta_l},
\qquad
\beta_l\beta_l^*=\frac{1-qN_l}{1-q}\prod_{1\leq r<s\leq3}(1-t_rt_sN_0)^{\delta_l},
\end{gather*}
preserving the ultralocality:
\begin{gather*}
\beta_l\beta_k=\beta_k\beta_l,
\qquad
\beta_l^*\beta_k^*=\beta_k^*\beta_l^*,
\qquad
\beta_l\beta_k^*=\beta_k^*\beta_l,
\qquad
\beta_l^*\beta_k=\beta_k\beta_l^*
\end{gather*}
if $l<k$.
The corresponding $q$-boson Hamiltonian $H$~\eqref{hamiltonian}, with the pertinent reduction of the
boundary potential $V(N_0,N_1)$~\eqref{boundary-potential} given by
\begin{gather*}
V(N_0,N_1)=\big(t_1+t_2+t_3-q^{-1}t_1t_2t_3N_0\big)\left(\frac{1-N_0}{1-q}\right)
+t_1t_2t_3N_0^2\left(\frac{1-N_1}{1-q}\right),
\end{gather*}
acts on $f$ in the $n$-particle subspace $\ell^2(\Lambda_n,\mathcal{N})$ via
\begin{gather*}
(H_{n}f)(\lambda)=\sum_{\substack{1\leq j\leq n\\
\text{\rm s.t.} \ \lambda+e_j\in\Lambda_n}}f(\lambda+e_j)[m_{\lambda_j}(\lambda)]\prod_{1\leq r<s\leq3}
\big(1-t_rt_sq^{m_0(\lambda)-1}\big)^{\delta_{\lambda_j}}
\\
\phantom{(H_{n}f)(\lambda)=}
{}+\sum_{\substack{1\leq j\leq n\\
\text{\rm s.t.} \ \lambda-e_j\in\Lambda_n}}f(\lambda-e_j)[m_{\lambda_j}(\lambda)]
\\
\phantom{(H_{n}f)(\lambda)=}
{}+f(\lambda)\left(\big(t_1+t_2+t_3-q^{-1}t_1t_2t_3q^{m_0(\lambda)}\big)[m_{0}
(\lambda)]+t_1t_2t_3q^{2m_0(\lambda)}[m_{1}(\lambda)]\right).
\end{gather*}

\subsection{Two-parameter reduction}

From the def\/ining orthogonality and the triangularity properties of
the hyperoctahedral Hall--Littlewood polynomials $\text{p}_\lambda$, $\lambda\in\Lambda_n$ detailed in
Section~\ref{orthogonality:sec},
it is read-of\/f that for $t_3,t_4\to 0$ these
polynomials reduce to Macdonald's Hall--Littlewood polynomials associated with the $BC$ type root
system~\cite[\S~10]{mac:orthogonal}.
This implies that they can be rewritten in terms of Macdonald's formula:
\begin{subequations}\label{msf}
\begin{gather}
\label{msfa}
\text{p}_\lambda(\xi)=\mathcal{N}_\lambda\sum_{w\in W}C(w\xi)e^{-i\langle\lambda,w\xi\rangle},
\end{gather}
with
\begin{gather}\label{msfb}
C(\xi)=\prod_{1\leq j<k\leq n}
\frac{\big(1-q e^{i(\xi_j-\xi_k)}\big)\big(1-q e^{i(\xi_j+\xi_k)}\big)}
{\big(1-e^{i(\xi_j-\xi_k)}\big)\big(1-e^{i(\xi_j+\xi_k)}\big)}
\prod_{1\leq j\leq n}\frac{\big(1-t_1e^{i\xi_j}\big)\big(1-t_2e^{i\xi_j}\big)}{1-e^{2i\xi_j}}
\end{gather}
and
\begin{gather}\label{msfc}
\mathcal{N}_\lambda=\frac{(1-q)^n}{(t_1t_2)_{m_0(\lambda)}\prod\limits_{l\geq0}(q)_{m_l(\lambda)}}.
\end{gather}
\end{subequations}
Notice in this connection that one does {\em not} directly retrieve Macdonald's
formula~\eqref{msf} by performing the limit $t_3,t_4\to 0$ in Venkateswaran's
formula~\eqref{HL-monic-normalization}.
Instead, the equivalence of the two formulas (for this specialization of the parameters) is not obvious and
rather {\em follows} from the fact that both expressions represent the same polynomials of the form in
equations~\eqref{hl} and~$\Delta$ given by equation~\eqref{measure} with
$t_3=t_4=0$~\cite{mac:orthogonal,ven:symmetric}.
Since the expansion coef\/f\/icients $C(w\xi)$ in equations~\eqref{msf} no longer depend on
(the number of nonzero parts of) $\lambda$, in the present situation the coordinate Bethe ansatz form of
the wave function~$\phi_\xi$~\eqref{wave-function} is seen to extend from the bulk sites (at~$\ell >0$) to
the boundary site (at~$\ell =0$).

The actions of the annihilation and creation operators~\eqref{an-op-cr-op} on
$f\in\ell^2(\Lambda_n,\mathcal{N})$ now reduce~to
\begin{subequations}\label{an-op-2-cr-op-2}
\begin{gather}
\label{an-op-2}
(\beta_lf)(\lambda)=f(\beta_l^*\lambda),
\qquad
 \lambda\in\Lambda_{n-1}
\end{gather}
with the convention that $\beta_lf=0$ if $n=0$, and
\begin{gather}
\label{cr-op-2}
(\beta_l^*f)(\lambda)=f(\beta_l\lambda)[m_l(\lambda)]\big(1-t_1t_2q^{m_0(\lambda)-1}\big)^{\delta_l},
\qquad
 \lambda\in\Lambda_{n+1}
\end{gather}
\end{subequations}
with the convention that $(\beta_l^*f)(\lambda )=0$ if $m_l(\lambda)=0$.
We thus arrive at an ultralocal two-parameter deformation of the $q$-boson f\/ield algebra at the boundary
site $l=0$ represented by~$\beta_l$,~$\beta_l^*$~\eqref{an-op-2-cr-op-2} and $N_l$~\eqref{num-op},
$l\in\mathbb{N}$:
\begin{gather*}
\beta_l N_k=q^{\delta_{l-k}}N_k\beta_l,
\qquad
\beta_l^*N_k=q^{-\delta_{l-k}}N_k\beta_l^*,
\\
\beta_l^*\beta_l=\frac{1-N_l}{1-q}\big(1-q^{-1}t_1t_2N_0\big)^{\delta_l},
\qquad
\beta_l\beta_l^*=\frac{1-qN_l}{1-q}(1-t_1t_2N_0)^{\delta_l},
\end{gather*}
and
\begin{gather*}
\beta_l\beta_k=\beta_k\beta_l,
\qquad
\beta_l^*\beta_k^*=\beta_k^*\beta_l^*,
\qquad
\beta_l\beta_k^*=\beta_k^*\beta_l,
\qquad
\beta_l^*\beta_k=\beta_k\beta_l^*
\end{gather*}
if $l<k$.
The corresponding $q$-boson Hamiltonian $H$~\eqref{hamiltonian}, with the reduction of the boundary
potential $V(N_0,N_1)$~\eqref{boundary-potential} given by
\begin{gather*}
V(N_0,N_1)=V(N_0):=\left(t_1+t_2\right)\left(\frac{1-N_0}{1-q}\right),
\end{gather*}
acts on $f$ in the $n$-particle subspace $\ell^2(\Lambda_n,\mathcal{N})$ via
\begin{gather}
(H_{n}f)(\lambda)
=\sum_{\substack{1\leq j\leq n\\\text{\rm s.t.} \ \lambda+e_j\in\Lambda_n}}
f(\lambda+e_j)[m_{\lambda_j}(\lambda)]\big(1-t_1t_2q^{m_0(\lambda)-1}\big)^{\delta_{\lambda_j}}
\nonumber
\\
\phantom{(H_{n}f)(\lambda)=}
{}+\sum_{\substack{1\leq j\leq n\\\text{\rm s.t.} \ \lambda-e_j\in\Lambda_n}}f(\lambda-e_j)[m_{\lambda_j}(\lambda)]
+f(\lambda)\left(t_1+t_2\right)[m_0(\lambda)].
\label{Hn-two}
\end{gather}
The latter semi-inf\/inite $q$-boson model with two-parameter boundary interactions was introduced and
studied in more detail in~\cite{die-ems:semi-infinite}.

\begin{remark}
When $q\to 0$ and $t_r\to 0$ ($r=1,\ldots,4$), the action of our $n$-particle Hamilto\-nian~$H_n$ on
$f:\Lambda_n\to\mathbb{C}$ reduces to that of a~discrete Laplacian
\begin{gather}
\label{Hn0}
(H_{n,0}f)(\lambda)=\sum_{\substack{1\leq j\leq n
\\
\text{\rm s.t.~} \lambda+e_j\in\Lambda_n}}f(\lambda+e_j)+\sum_{\substack{1\leq j\leq n
\\
\text{\rm s.t.~} \lambda-e_j\in\Lambda_n}}f(\lambda-e_j)
\nonumber
\end{gather}
modeling a~system of $n$ impenetrable bosons on $\mathbb{N}$.
In~\cite[Section~5]{die-ems:semi-infinite} it was shown that the large-times asymptotics of the
$q$-boson dynamics generated by $H_n$~\eqref{Hn-two} is related to the impenetrable boson dynamics of
$H_{n,0}$~\eqref{Hn0} via an $n$-particle scattering matrix of the form
\begin{subequations}
\begin{gather}
{\mathcal S}(\xi)=\prod_{1\leq j<k\leq n}s(\xi_j-\xi_k)s(\xi_j+\xi_k)\prod_{1\leq j\leq n}s_0(\xi_j),
\end{gather}
with
\begin{gather}
\label{s-s0}
s(x)=\frac{1-qe^{-ix}}{1-qe^{ix}}
\qquad
\text{and}
\qquad
s_0(x)=\frac{(1-t_1e^{-ix})(1-t_2e^{-ix})}{(1-t_1e^{ix})(1-t_2e^{ix})}.
\end{gather}
\end{subequations}
The discussion in~\cite[Section~5]{die-ems:semi-infinite} applies verbatim to our more general
Hamiltonian $H_n$ from Proposition~\ref{Hn:prp} with $V(N_0,N_1)$ given by
equation~\eqref{boundary-potential}, upon replacing $s_0(x)$~\eqref{s-s0} by
\begin{gather*}
s_0(x)=\prod_{r=1}^4\frac{1-t_re^{-ix}}{1-t_re^{ix}}.
\end{gather*}
This reveals that the $n$-particle scattering matrix of the model factorizes in two-particle bulk
scattering matrices $s(\cdot)$ governed by a~coupling parameter $q$ and one-particle boundary scattering
matrices $s_0(\cdot)$ governed by coupling parameters $t_1,\ldots,t_4$.
\end{remark}

\subsection*{Acknowledgments} We are grateful to Alexei Borodin and Ivan Corwin for helpful email
exchanges and thank the referees for their constructive comments.
This work was supported in part by the {\em Fondo Nacional de Desarrollo Cient\'{\i}fico y Tecnol\'ogico
$($FONDECYT$)$} Grants \# 1130226 and \#~11100315, and by the {\em Anillo ACT56 `Reticulados y Simetr\'{\i}as'}
f\/inanced by the {\em Comisi\'on Nacional de Investigaci\'on Cient\'{\i}fica y Tecnol\'ogica $($CONICYT$)$}.

\pdfbookmark[1]{References}{ref}
\LastPageEnding


\begin{thebibliography}{99}
\footnotesize\itemsep=0pt

\bibitem{bog-ize-kit:correlation}
Bogoliubov N.M., Izergin A.G., Kitanine N.A., Correlation functions for a
  strongly correlated boson system, \href{http://dx.doi.org/10.1016/S0550-3213(98)00038-8}{\textit{Nuclear Phys.~B}} \textbf{516}
  (1998), 501--528, \href{http://arxiv.org/abs/solv-int/9710002}{solv-int/9710002}.

\bibitem{bor-cor-pet-sas:spectral}
Borodin A., Corwin I., Petrov L., Sasamoto T., Spectral theory for the
  $q$-boson particle system, \href{http://arxiv.org/abs/1308.3475}{arXiv:1308.3475}.

\bibitem{die:properties}
van Diejen J.F., Properties of some families of hypergeometric orthogonal
  polynomials in several variables, \href{http://dx.doi.org/10.1090/S0002-9947-99-02000-0}{\textit{Trans. Amer. Math. Soc.}}
  \textbf{351} (1999), 233--270, \href{http://arxiv.org/abs/q-alg/9604004}{q-alg/9604004}.

\bibitem{die-ems:diagonalization}
van Diejen J.F., Emsiz E., Diagonalization of the inf\/inite $q$-boson system,
  \href{http://arxiv.org/abs/1308.2237}{arXiv:1308.2237}.

\bibitem{die-ems:semi-infinite}
van Diejen J.F., Emsiz E., The semi-inf\/inite $q$-boson system with boundary
  interaction, \href{http://dx.doi.org/10.1007/s11005-013-0657-y}{\textit{Lett. Math. Phys.}}, {t}o appear, \href{http://arxiv.org/abs/1308.2242}{arXiv:1308.2242}.

\bibitem{kli-sch:quantum}
Klimyk A., Schm{\"u}dgen K., Quantum groups and their representations, \textit{Texts
  and Monographs in Physics}, Springer-Verlag, Berlin, 1997.

\bibitem{koo:askey-wilson}
Koornwinder T.H., Askey--{W}ilson polynomials for root systems of type {$BC$},
  in Hypergeometric Functions on Domains of Positivity, {J}ack Polynomials, and
  Applications ({T}ampa, {FL}, 1991), \href{http://dx.doi.org/10.1090/conm/138/1199128}{\textit{Contemp. Math.}}, Vol.~138, Amer.
  Math. Soc., Providence, RI, 1992, 189--204.

\bibitem{kor:cylindric}
Korf\/f C., Cylindric versions of specialised {M}acdonald functions and a
  deformed {V}erlinde algebra, \href{http://dx.doi.org/10.1007/s00220-012-1630-9}{\textit{Comm. Math. Phys.}} \textbf{318} (2013),
  173--246, \href{http://arxiv.org/abs/1110.6356}{arXiv:1110.6356}.

\bibitem{mac:orthogonal}
Macdonald I.G., Orthogonal polynomials associated with root systems,
  \textit{S\'em. Lothar. Combin.} \textbf{45} (2000), Art.~B45a, 40~pages,
  \href{http://arxiv.org/abs/math.QA/0011046}{math.QA/0011046}.

\bibitem{mac:affine}
Macdonald I.G., Af\/f\/ine {H}ecke algebras and orthogonal polynomials,
  \href{http://dx.doi.org/10.1017/CBO9780511542824}{\textit{Cambridge Tracts in Mathematics}}, Vol.~157, Cambridge University
  Press, Cambridge, 2003.

\bibitem{maj:foundations}
Majid S., Foundations of quantum group theory, \href{http://dx.doi.org/10.1017/CBO9780511613104}{Cambridge University Press},
  Cambridge, 1995.

\bibitem{pov:integrability}
Povolotsky A.M., On integrability of zero-range chipping models with factorized
  steady state, \href{http://dx.doi.org/10.1088/1751-8113/46/46/465205}{\textit{J.~Phys.~A: Math. Theor.}} \textbf{46} (2013), 465205,
  25~pages, \href{http://arxiv.org/abs/1308.3250}{arXiv:1308.3250}.

\bibitem{sas-was:exact}
Sasamoto T., Wadati M., Exact results for one-dimensional totally asymmetric
  dif\/fusion models, \href{http://dx.doi.org/10.1088/0305-4470/31/28/019}{\textit{J.~Phys.~A: Math. Gen.}} \textbf{31} (1998),
  6057--6071.

\bibitem{tak:discrete}
Takeyama Y., A discrete analogue of periodic delta Bose gas and af\/f\/ine Hecke
  algebra, \href{http://arxiv.org/abs/1209.2758}{arXiv:1209.2758}.

\bibitem{tsi:quantum}
Tsilevich N.V., The quantum inverse scattering problem method for the
  {$q$}-boson model and symmetric functions, \href{http://dx.doi.org/10.1007/s10688-006-0032-1}{\textit{Funct. Anal. Appl.}}
  \textbf{40} (2006), 207--217, \href{http://arxiv.org/abs/math-ph/0510073}{math-ph/0510073}.

\bibitem{ven:symmetric}
Venkateswaran V., Symmetric and nonsymmetric Hall--Littlewood polynomials of
  type~$BC$, \href{http://arxiv.org/abs/1209.2933}{arXiv:1209.2933}.

\end{thebibliography}
\end{document}